\documentclass[prd,preprintnumbers,amsmath,amssymb,floatfix,showpacs]{revtex4}

\usepackage{graphicx}
\usepackage{dcolumn}
\usepackage{bm}
\usepackage{amssymb}
\usepackage{epsfig}
\usepackage{color}

\newcommand{\ba}{\begin{eqnarray}}
\newcommand{\ea}{\end{eqnarray}}
\newcommand{\be}{\begin{equation}}
\newcommand{\ee}{\end{equation}}
\newcommand{\bdisplay}{\begin{displaymath}}
\newcommand{\edisplay}{\end{displaymath}}

\newcommand{\eq}[1]{Eq.\,(\ref{#1})}
\newcommand{\fig}[1]{Fig.\,\ref{#1}}

\newcommand{\alphabold}{\mbox{\small\boldmath $\alpha$}}

\newcommand{\delchisqW}{\Delta \chi^2_i(W_i;\alphabold)}

\newcommand{\delchi}{\Delta \chi^2_i}

\newcommand{\delchimax}{{\delchi}_{\rm max}}

\begin{document}

\title{Comprehensive fits to high energy data for $\sigma$, $\rho$, and $B$ and the asymptotic black-disk limit}

\author{Martin~M.~Block}
\email{mblock@northwestern.edu}
\affiliation{Department of Physics and Astronomy, Northwestern University,
Evanston, IL 60208}
\author{Loyal Durand}
\email{ldurand@hep.wisc.edu}
\altaffiliation{Mailing address: 415 Pearl Ct., Aspen, CO 81611}
\affiliation{Department of Physics, University of Wisconsin, Madison, WI 53706}
\author{Phuoc Ha}
\email{pdha@towson.edu}
\affiliation{Department of Physics, Astronomy and Geosciences, Towson University, Towson, MD 21252}
\author{Francis Halzen}
\email{francis.halzen@icecube.wisc.edu}
\affiliation{Wisconsin IceCube Particle Astrophysics Center and Department
of Physics, University of Wisconsin-Madison,  Madison,  Wisconsin 53706}

\begin{abstract}

We demonstrate that the entirety of the data on proton--proton and antiproton--proton  forward scattering between 6 GeV and 57 TeV center-of-mass energy is sufficient to show that $\sigma_{\rm elas}/\sigma_{\rm tot} \rightarrow 1/2$, and that  $8\pi B/\sigma_{\rm tot}\rightarrow 1$ at very high energies, where $B$ the forward slope parameter  for the differential elastic scattering cross sections. The relations demonstrate convincingly that the asymptotic $pp$ and $\bar{p}p$ scattering amplitudes approach those of scattering from a black disk.  This result obviously has implications for any new physics that modifies the forward scattering amplitudes.

\end{abstract}


\date{\today}

\maketitle

%%%%%%%%%%%%% SEC. INTRODUCTION %%%%%%%%%%

\section{Introduction \label{sec:introduction}}

Proton--proton ($pp$) and antiproton--proton ($\bar{p}p$) scattering have been studied for many decades. A persistent question since the advent of high-energy accelerators has concerned the behavior of the cross sections at very high energies. They are bounded theoretically to increase no more rapidly than $\ln^2{s}$, the Froissart bound \cite{froissart,martin1,martin4}, where $s=W^2$ is the square of the total energy in the center-of-mass system. Block and Halzen \cite{blockhalzenfit} and Igi and Ishida \cite{Igi,Igi2} showed convincingly that the $\ln^2{s}$ behavior in fact held for the $pp$ and $\bar{p}p$ cross sections measured up to Tevatron energies, with this behavior leading to successful predictions for the cross sections at the Large Hadron Collider (LHC). For a review, see \cite{blockrev}.

Block and Halzen \cite{blockhalzen,blockhalzen2} and Schegelsky and Ryskin \cite{ryskin}  also presented tentative evidence that the  $pp$ and $\bar{p}p$ scattering amplitudes may asymptotically approach those for scattering from a completely absorptive or ``black'' disk---the ``black-disk'' limit---at ultra-high energies, but the results of those analyses were not definitive. This result, and the common assumption that hadronic scattering is dominated at high energies by the interactions between gluons in the two hadrons, together imply that all hadron-hadron cross sections should approach a common black-disk limit as $s\rightarrow\infty$, a very interesting result.

In the present paper, we present the results of a comprehensive analysis  of the forward $pp$ and $\bar{p}p$ scattering data for center-of--mass energies from 6 GeV to 57 TeV. We discuss various constraints on the cross sections which are essential in tying down the parametrizations of the low-energy cross sections, and present a fit to the data on $\sigma_{\rm tot}^{pp(\bar{p}p)}$, $\sigma_{\rm elas}^{pp(\bar{p}p)}$, and $\sigma_{\rm inel}^{pp(\bar{p}p)}$, the forward slope parameters $B^{pp}$ and $B^{\bar{p}p}$, and the ratios of the real to imaginary parts of the forward scattering amplitudes $\rho^{pp}$ and $\rho^{\bar{p}p}$, using parametrizations which reflect the established $\ln^2{s}$ behavior of the cross sections at high energies.

We find that the fit to the entirety of the data gives convincing evidence that the  $pp$ and $\bar{p}p$ scattering amplitudes approach the black-disk limit at very high energies. We use this result to obtain a final, essentially identical, fit with the black-disk constraints $\sigma_{\rm elas}/\sigma_{\rm tot} \rightarrow 1/2$, and  $8\pi B/\sigma_{\rm tot}\rightarrow 1$ imposed from the outset. The results give  predictions with only small uncertainties for the cross sections, $\rho$, and $B$ at the higher energies which may become accessible in the future. The present results agree well with the predictions of earlier fits

%%%%%%%%%%%%%%%%%%%%%%%%%
%%%%%%%%%%%%%%%%%%%%%%%%%

\section{Parametrizations and constraints \label{sec:input}}

%%%%%%%%%%%%%%%%%%%%%%%%%

\subsection{Parameterization of the cross sections, the real-to-imaginary ratio $\rho$, and the slope parameter $B$ \label{subset:parametrizations}}

We will be concerned here with global fits to the high-energy total, elastic, and inelastic $pp$ and $\bar{p}p$ scattering cross sections, the ratios $\rho={\rm Re}f(s,0)/{\rm Im}f(s,0)$ of the real to the imaginary parts of the forward elastic scattering amplitudes $f(s,t)$, and the forward slope parameter $B=d(\ln\sigma(s,t))/dt\big|_{t=0}$ for the differential cross sections $d\sigma/dt$. We will use the parametrizations of $\sigma_{\rm tot}$ and $\rho$ introduced by  Block and Cahn \cite{blockcahn} and used by Block and Halzen \cite{blockhalzenfit,blockrev}  in their earlier fit to the $pp$ and $\bar{p}p$ data up to a center-of-mass energy $W=\sqrt{s}=1800$ GeV. That fit was excellent and gave successful predictions of the more recent, higher energy data from the Large Hadron Collider (LHC) and cosmic ray experiments \cite{blockhalzen,blockhalzen2}.

The Block-Cahn analysis assumed a $\ln^2s$ bound on the growth of the cross sections at high energy as implied by the Froissart bound \cite{froissart,martin1,martin4} and parametrized $\sigma_{\rm tot}^{pp}$ and $\sigma_{\rm tot}^{\bar{p}p}$ as quadratic expressions in the $s-$dependent variable $\nu/m =(s-2m^2)/2m^2$ with additional falling Regge-like terms important at lower energies. The phase of the scattering amplitude at high energies and the corresponding expression for $\rho$ then followed from the constraints imposed by analyticity and crossing symmetry under the transformation $\nu\rightarrow-\nu$ \cite{blockcahn,blockrev}.

We will extend the parametrizations here to the elastic and inelastic cross sections and the $B$ parameter, with
\ba
\label{sigma0}
\sigma^{\rm 0}(\nu) &=& c_0+c_1\ln\left(\frac{\nu}{m}\right)+c_2\ln^2\left(\frac{\nu}{m}\right)+\beta\left(\frac{\nu}{m}\right)^{\mu-1} \, , \\
\label{sigma_tot}
\sigma_{\rm tot}^{\pm}(\nu) &=& \sigma^{\rm 0}(\nu) \pm\delta \left(\frac{\nu}{m}\right)^{\alpha-1}, \\
\label{sigma_el}
\sigma_{\rm elas}^{\pm}(\nu) &=& b_0+b_1\ln\left(\frac{\nu}{m}\right)+b_2\ln^2\left(\frac{\nu}{m}\right)+\beta_e\left(\frac{\nu}{m}\right)^{\mu-1} \pm\delta_e \left(\frac{\nu}{m}\right)^{\alpha-1} \, , \\
\label{rho}
\rho^{\pm} &=& \frac{1}{\sigma_{\rm tot}^{\pm}(\nu)}\left[\frac{\pi}{2}c_1+\pi c_2\ln\left(\frac{\nu}{m}\right) - \beta \cot\left(\frac{\pi\mu}{2}\right) \left(\frac{\nu}{m}\right)^{\mu-1} +\frac{4\pi}{\nu}f_+(0)
 \pm \delta\tan\left(\frac{\pi\alpha}{2}\right)\left(\frac{\nu}{m}\right)^{\alpha-1}\right], \\
 \label{B}
B^{\pm}(\nu) &=& a_0+a_1\ln\left(\frac{\nu}{m}\right)+a_2\ln^2\left(\frac{\nu}{m}\right)+\beta_{B}\left(\frac{\nu}{m}\right)^{\mu-1} \pm\delta_B \left(\frac{\nu}{m}\right)^{\alpha-1},
\ea
where the upper and lower signs are for $pp$ and $\bar{p}p$ scattering, respectively. Here $\nu$ is the laboratory energy of the incident particle, with $2m\nu=s-2m^2=W^2-2m^2$ where $W$ is the center-of-mass energy and $m$ is the proton mass. The inelastic cross sections are given by the differences between the total and elastic cross sections, $\sigma_{\rm inel}^{\pm} = \sigma_{\rm tot}^{\pm} - \sigma_{\rm elas}^{\pm}$. They are therefore parametrized simply as the differences of the expressions in Eqs.\ (\ref{sigma_tot}) and (\ref{sigma_el}); no new parameters appear.

It is not obvious that the very simple parametrizations above should be adequate to describe the cross sections, $\rho$, and $B$ over the entire energy range we will consider. It is also not clear that the  coefficients of these terms can be determined well enough from fits in the extant energy range to extrapolate properly into the ultra-high energy region where the $\ln^2(\nu/m)$ terms become dominant.  We have studied these questions quantitatively using a detailed eikonal model which provides a very good description of the data from 4 GeV to 57 TeV \cite{bdhh_eikonal}. In that analysis, we used the expressions above to fit  ``data'' for the cross sections, $\rho$, and $B$ derived from the eikonal model. The fits  are excellent, with errors typically smaller than the real experimental uncertainties, and those fits over the ``experimental'' region continue to hold to ultra-high energies. Small correction terms would certainly be present analytically in the expressions in Eqs.\ (\ref{sigma0})--(\ref{B}), but these are clearly unimportant in the fitting and extrapolation.

We emphasize also that the presence of the $\ln^2(\nu/m)$ terms in the parametrizations  is not connected directly with the Froissart bound: these terms are consistent with the bound, but follow in the eikonal model  from the power-law growth of the imaginary part of the eikonal function coupled with its exponentially bounded behavior in impact parameter space. This leads to  a effective radius of interaction between the nucleons  that grows logarithmically with increasing energy, and within which the scattering is nearly completely absorptive. As a result, the  scattering approaches the ``black-disk'' limit at very high energies, with consequences we discuss below. Finally, as noted in \cite{bdhh_eikonal}, the  coefficients of the $\ln^2(\nu/m)$ terms depend on properties of the eikonal function that are not well determined. We therefore argued that the best extrapolations of cross sections and other parameters to ultra-high energies are those based on direct fits to the data using the parametrization above. We carry out those fits here.

 We turn next to a discussion of the known constraints on the parameters in Eqs.\ (\ref{sigma0})--(\ref{B}).

%%%%%%%%%%%%%%%%%%%%%%%%%
%%%%%%%%%%%%%%%%%%%%%%%%%

\subsection{Constraints \label{subsec:constraints}}

\subsubsection{Low-energy constraints \label{subsubsec:LEconstraints}}

There are nominally 18 parameters ($a_0, a_1, a_2, b_0, b_1, b_2, c_0, c_1, c_2, \beta, \beta_e, \beta_B, \delta, \delta_e, \delta_B, \alpha$, $\mu$, and $f_+(0)$) in the model, but these are not all independent and must satisfy certain constraints. When these are  imposed, we will end up with only 12 independent parameters in our final fit.

Both the ``analyticity constraints'' of Block and Halzen, derived in \cite{block_analytic} and discussed in detail in \cite{blockhalzenfit}, and the finite energy sum rule (FESR2) of Igi and Ishida \cite{Igi,blockrev}, impose constraints on the parameters. The first requires that the fits reproduce the values of the total cross sections at a transition point $\nu_0$ far enough above the resonance region that the high-energy parametrizations may be expected to hold, but where the cross sections can still be evaluated accurately using the dense low-energy data. The second approach obtains equivalent results through a matching of the FESR integrals at $\nu_0$. Following \cite{blockhalzenfit}, we take $\nu_0=7.59$ GeV corresponding to $W=\sqrt{s}=4$ GeV. Their low-energy analysis gives $\sigma_{\rm tot}^{pp}=40.18$ mb, $\sigma_{\rm tot}^{\bar{p}p}=56.99$ mb.

 In the case of the  crossing-even combination of cross sections $\sigma^0=\left(\sigma_{\rm tot}^{pp}+\sigma_{\rm tot}^{\bar{p}p}\right)/2=\left(\sigma_{\rm tot}^++\sigma_{\rm tot}^-\right)/2$ this matching gives the constraint
\be
\label{analyticity_const}
c_0+c_1\ln(\nu_0/m)+c_2\ln^2(\nu_0/m)+\beta(\nu_0/m)^{\mu-1}=\sigma^0(\nu_0)=48.58\ {\rm mb}.
\ee
An essentially equivalent result numerically follows from the finite-energy sum rules of Igi and Ishida \cite{Igi,Igi2,blockrev} relating the low- and high-energy regions \cite{blockrev}.

A second  constraint holds for the crossing-odd combination of cross sections $\Delta\sigma=\left(\sigma_{\rm tot}^+-\sigma_{\rm tot}^-\right)/2$. Matching the theoretical and experimental results, we find that
\be
\label{analyticity_const2}
\delta\left(\nu_0/m\right)^{\alpha-1}=\Delta\sigma(\nu_0)= -8.405\ {\rm mb}.
\ee

Two further analyticity constraints hold if one matches the derivatives of the cross sections with respect to $\nu/m$ to their experimental values at $\nu_0$  \cite{blockhalzenfit}. We will not use these because they are less reliable numerically and are more sensitive than the cross sections themselves to small deviations of the high-energy expressions in Eqs.\ (\ref{sigma0}) and (\ref{sigma_tot}) from the actual cross sections at the rather low matching energy of 4 GeV.

A rather subtle constraint holds for the coefficients $\beta,\,\beta_e,\,\delta,\,\delta_e$ of the Regge-like terms. These cannot be entirely independent since a descending power-law term in the eikonal function in a general impact-parameter representation of the scattering amplitudes $f^{\pm}(s,t)$ affects $\sigma_{\rm elas}^{\pm}$ and $\sigma_{\rm inel}^{\pm}$ as well as $\sigma_{\rm tot}^{\pm}$.  We have investigated these aspects of the scattering using our detailed eikonal model for $pp$ and $\bar{p}p$ scattering \cite{bdhh_eikonal}, which gives an accurate description of the data over the region where the Regge-like effects are important.

The cross sections are described in the eikonal model in terms of the integrals
\ba
\label{sigma_total}
\sigma_{\rm tot}(s) &=&4\pi {\rm Im} f(s,0) = 4\pi \int_0^\infty db\, b \left (1-\cos{\chi_R}\,e^{-\chi_I}\right), \\
\label{sigma_elas}
\sigma_{\rm elas} &=& 2\pi\int_0^\infty db\, b \left |1-e^{i\chi}\right |^2 = 2\pi\int_0^\infty db\, b \left(1-2\cos{\chi_R}\, e^{-\chi_I}+e^{-2\chi_I}\right), \\
\label{sigma_inel}
\sigma_{\rm inel}(s) &=& \sigma_{\rm tot}-\sigma_{\rm elas} = 2\pi\int_0^\infty db\, b \left (1-e^{-2\chi_I}\right),
\ea
where $\chi=\chi_R+i\chi_I$ is the complex eikonal function written in terms of crossing-even and crossing-odd parts.

Writing $\chi$ as $\chi=\chi^0+\chi^{\rm Regge}$, we can isolate the contributions of the Regge-like terms to the crossing-even and crossing-odd cross sections $\sigma^0(\nu)$ and $\Delta\sigma(\nu)$ by subtracting the expression for the cross section for $\chi^{\rm Regge}=0$ from the full result.  The effect of the factor $\cos{\chi_R}$ in \eq{sigma_tot} is small enough that we can neglect it for this purpose. If we do so,  the contribution of the crossing-even Regge term to the total cross section $\sigma^0(\nu)$ is given by the expression
\be
\label{Regge_tot}
4\pi \int_0^\infty db\,b\cosh{\!\left(\chi_I^{\rm Regge, odd}\right)} \,e^{-\chi_I^{\rm 0,even}}\left(1-e^{-\chi_I^{\rm Regge,even}}\right).
\ee
We note that the contribution of $\chi_I^{\rm Regge, odd}$ through the $\cosh$ function is second order in that quantity and can be dropped without significant loss of accuracy. Similar expressions hold for the other cross sections.

Despite the somewhat different effects of the energy-dependent eikonalization in the different cross sections, we find that the input power in a Regge-like term  $(m/\nu)^\gamma$ in the eikonal function  $\chi^{\rm Regge}$ is reproduced to a percent or better in  output power-law fits to the various  integrals over the energy interval 6--1000 GeV, where those outputs are to be identified with the Regge-like terms in Eqs.\ (\ref{sigma0})--(\ref{B}). The powers are therefore stable across the expressions in Eqs.\ (\ref{sigma0})--(\ref{B}), as assumed.

Importantly, we find that the ratios of the crossing-even and crossing-odd Regge-like contributions to $\sigma_{\rm inel}$  to the corresponding contributions  to $\sigma_{\rm tot}$ vary only slowly over the most important important energy range, 6 to 100 GeV (and beyond), with the even ratio in the range 0.684--0.657 and the odd ratio in the range 0.802--0.787. Averages weighted by the even- and odd cross sections give ratios 0.678 and 0.797.

Converting these results on the Regge-like terms to the elastic and total cross sections \eq{sigma_tot} and \eq{sigma_el}, we find that
\be
\label{ratio_const}
\beta_e=0.302\,\beta, \quad  \delta_e=0.203\,\delta
\ee
 as averaged over the interval 6--100 GeV, with only very small variations from these values. These relations give our  new, and not-very-obvious, constraints on the $\beta$ and $\delta$ parameters in  Eqs.\ (\ref{sigma0}) and (\ref{sigma_el}). The smallness of the elastic-to-total ratios is easily understood: the Regge-like terms enter the elastic cross section in \eq{sigma_el} only in second order in $\chi^{\rm Regge}$, but appear to first order in $\sigma_{\rm tot}$ and $\sigma_{\rm inel}$.

With the imposition of the 4 low-energy constraints in Eqs.\ (\ref{analyticity_const}), (\ref{analyticity_const2}), and (\ref{ratio_const}), 14 parameters are left to fit all data using the parametrizations introduced above.   These constraints are quite important: the results anchor the total cross sections accurately at the starting energies and in the Regge region, removing extra parameters which can otherwise mix with and affect the values of the high-energy parameters of primary interest. We note that only 9 of the remaining parameters appear in the expressions for the total, elastic, and inelastic cross sections and $\rho$; the remaining 5 are in the expression for $B$.

%%%%%%%%%%%%%%%%%%%%%%%%%

 \subsubsection{High-energy constraints \label{subsubsec:HEconstraints}}

 As noted above, we expect the $pp$ and $\bar{p}p$ scattering amplitudes to approach the black-disk limit at ultra-high energies, with the scattering amplitudes approaching those for scattering from a completely absorbing disk with a radius $R$ which increases logarithmically with energy.  In that limit,  $\chi_R\rightarrow 0$ while $e^{-\chi_I}$ vanishes for impact parameters $0\leq b\leq R$ and is equal to 1 for $b>R$. As a result, from \eq{sigma_total}, $\sigma_{\rm tot}\rightarrow 2\pi R^2$  up to edge effects of order $R$ \cite{bdhh_eikonal}, while from \eq{sigma_elas}, $\sigma_{\rm elas}\rightarrow \pi R^2$, also up to edge effects, and $\sigma_{\rm elas}/\sigma_{\rm tot}\rightarrow 1/2$.

 The real part of the forward scattering amplitude $f(s,0)$ is associated at high energies with peripheral scattering outside the region of strong absorption and, as an edge effect, is proportional to $R$ for finite-range forces. It therefore decreases as $1/R$ relative to the imaginary part which is proportional to $\sigma_{\rm tot}\propto R^2$, and $\rho\propto 1/R\propto 1/\ln{W} \rightarrow 0$ at high energies.

 Finally, for ${\rm Re}\,f(s,0)\ll {\rm Im}\,f(s,0)$, the slope parameter $B$ can be written as \cite{bdhh_eikonal}
 \be
 \label{B_formula}
 B = \frac{1}{2}\int_0^\infty db\,b^3\left(1-e^{-\chi_I}\right)\bigg/\int_0^\infty db\,b\left(1-e^{-\chi_I}\right).
 \ee
 With the conditions above, the integrals can be evaluated simply in the black-disk limit, and we find that
 \be
 \label{B_limit}
 B \rightarrow R^2/4=\sigma_{\rm tot}/8\pi.
 \ee

 The same result for $B$ can be derived less rigorously if it is assumed that the differential scattering cross section is purely  exponential in $t$, with $d\sigma_{\rm elas}/dt=\pi\left| f(s,0)\right|^2 e^{Bt}$. Integrating over $t$ from $-\infty$ to 0, then using the the relation $\left|f(s,0)\right|^2=16\pi^2\left(1+\rho^2\right) \sigma_{\rm tot}^2$ and rearranging, we find that  \cite{blockcahn} $B=\sigma_{\rm tot}^2\left(1+\rho^2\right)\big/ 16\pi^2\sigma_{\rm elas}$, or, with $\rho\rightarrow 0$ and $\sigma_{\rm elas}/\sigma_{\rm tot}\rightarrow 1/2$, $B\rightarrow\sigma_{\rm tot}/8\pi$.

 It is an important question as to whether there is evidence of an approach to the black-disk limit in present data. If so, it is reasonable to impose the black disk constraints $\sigma_{\rm inel}/\sigma_{\rm tot}\rightarrow 1/2$ and $B\rightarrow \sigma_{\rm tot}/8\pi$ in a final fit to the data. This leads in the parametrization above to the constraints
 \be
 \label{HEconstraints}
 b_2=c_2/2,\qquad a_2=c_2/0.3894 \times 8 \pi
 \ee
where the numerical factor arises from the conversion of $c_2$ in mb to units of GeV$^{-2}$. This leaves 12 free parameters.

The approach to the black-disk limit was investigated for $\sigma_{\rm tot}$ and $\sigma_{\rm inel}$ in \cite{blockhalzen2} using a hybrid approach in which the parametrization for $\sigma_{\rm inel}$ was determined from that for $\sigma_{\rm tot}$ by multiplying the latter by the ratio  $\sigma_{\rm inel}/\sigma_{\rm tot}$ found in an earlier eikonal model \cite{blockaspen} and fitting the result to an expression of the form in \eq{sigma_tot}. The result agreed very well with the measured high-energy inelastic cross sections. The ratio of the coefficients of the $\ln^2(\nu/m)$ terms gave a value 0.509$\pm$0.021 in agreement with the expectation 1/2 for black-disk scattering, and was interpreted as evidence for this limit. This result can be questioned because the ratio $\sigma_{\rm inel}/\sigma_{\rm tot}$ has the asymptotic value 1/2 automatically in the eikonal model used to get the parametrization for $\sigma_{\rm inel}$ from that for $\sigma_{\rm tot}$. However, the excellent agreement of the predicted and measured inelastic scattering cross sections suggests that the same ratio should be found in a free fit to the data using the parametrization which follows from Eqs.\ (\ref{sigma_tot}) and (\ref{sigma_el}). We will examine this in the next section.

The asymptotic behavior of $B$ was studied by Schegelsky and Ryskin \cite{ryskin} who used a simple $a+b\ln^2(s/s_0)$ form with $s_0=1$ GeV$^2$ to fit  the high-energy data. The coefficient in their result, equivalent to $a_2= 0.0286 \pm 0.0005$  GeV$^{-2}$ in \eq{B}, and the relation in \eq{HEconstraints} predicted the value $c_2= 0.294 \pm 0.005$ mb for the leading coefficient  in $\sigma_{\rm tot}$, closely matching the value $c_2=0.2817 \pm 0.0064$ mb found in the analysis of \cite{blockhalzenfit}. This is again  evidence for the expected black disk behavior of the scattering at high energies. We note, however, that the fit in \cite{ryskin} is not tied down at low energies, with the result that those authors had to drop a more flexible parametrization to get their final result, even then with a $\chi^2$ per degree for freedom of 1.5, not a remarkably good fit. We will reexamine the fit to $B$ in the following section.

%%%%%%%%%%%%%%%%%%%%%%%
%%%%%%%%%%%%%%%%%%%%%%%

\section{Fits to high energy proton - proton and antiproton-proton data \label{sec:fit}}

%%%%%%%%%%%%%%%%%%%%%%%

\subsection{Data and method of fitting \label{subsec:fit_method}}

The data we will use in our analysis consists of results on $\sigma_{\rm tot}$ for $W\geq 6$ GeV,  $\sigma_{\rm inel}$ for $W\geq 540$ GeV,  $\sigma_{\rm elas}$ for $W\geq 30$ GeV, and  $\rho$  and $B$ for $W\geq 10$ GeV. The energy ranges for $\sigma_{\rm tot}$, $\sigma_{\rm inel}$, and $\rho$  are the same as used in the Block-Halzen fits \cite{blockhalzenfit,blockrev,blockhalzen}, but we include  the newer data  at very high energies from the LHC \cite{totem2011,totem2013,totem2013_2} and the Auger \cite{POAp-air} and HiRes \cite{HiRes} collaborations. As noted, we include the extensive data on $\sigma_{\rm elas}$ and $B$ in our fits; these quantities have not been used before in fits of this type.The data on $\sigma_{\rm elas}$ can be extended to 10 GeV or below without changing the final results significantly, but the data are somewhat less accurate in that region, and we prefer to emphasize the higher energies given our focus on the behavior of the cross sections and $B$ at ultra-high energies.

We used the sieve algorithm \cite{sieve,blockhalzenfit} to identify  outlying points and remove them from the data set used in the final fits. There are two underlying assumptions in this procedure. We assume, first, that the parametrization used in the fit, with the parameter set $\alphabold=\{a_0, a_1,\ldots,f_+(0)\}$, can give a good description of theory, a point checked theoretically in  \cite{bdhh_eikonal} for the present case. Second, we assume that  the complete data set consists mostly of datum points which have a normal Gaussian distribution with respect to the actual theoretical distribution, plus some outlying points which have a much broader distribution than reflected in their quoted (Gaussian) uncertainties, the result of unknown experimental problems. These outlying points can unduly influence a $\chi^2$ fit based on Gaussian statistics, but have much-reduced impact in a fit based on a broader statistical distribution.

The sieve procedure is based on a Lorentzian probability distribution adjusted to give results that agree very well with those from a Gaussian distribution in the absence of outliers, but which still eliminates the latter efficiently when they are present. The details of the analysis are given in the appendix to \cite{sieve}.

We first make a  fit to the complete data set by minimizing $\Lambda_0^2$, the Lorentzian squared with respect to the parameter set $\alphabold$ in the fit function over the datum points $y_i$ at the set {\boldmath{$W$}} of center-of-mass energies $W_i$ at which the observations are made,
\be
\label{lambda0^2}
\Lambda_0^2({\alphabold},\mbox{\boldmath{$W$}})=\sum_{i=1}^N\ln\left[1+0.179 \delchisqW\right].
\ee
Here $\delchisqW=\left[y_i-y_i(W_i,\alphabold)\right]^2/\sigma_i(W)$ where $y_i$ is the  value of the quantity of interest measured at energy $W_i$, $y_i(W_i,\alphabold)$ is the theoretical value of that quantity  for the parameters $\alphabold$, and $\sigma_i$ is the experimental error. Because of the intrinsically long tails of the Lorentzian distribution, this fit should be robust in the sense that points that lie far from the fitted distribution are accorded relatively little weight in the fitting, and do not influence the fit unduly.

We next eliminate datum points for which $\delchisqW$ is ``too large,'' with a value larger than a chosen $\Delta_{\rm max}$, taken here as $\Delta_{\rm max}=6$ \cite{sieve}. These points lie well away from the theoretical fit and are presumed to be outliers relative to the ``good'' Gaussian-distributed data. We then make a conventional Gaussian $\chi^2$ fit to the remaining points. If our assumptions about the nature of the distribution are correct, the parameters $\alphabold$ should not change significantly in this second fit, and the points identified as outliers should not change relative to the fit except possibly for those on the boundary with $\delchisqW\approx\Delta_{\rm max}$.

We note that 98.6\% (99.7\%) of the points in a normal Gaussian distribution would survive cuts with $\Delta_{\rm max}=6$ (9). However, the normal points eliminated would contribute significantly to the Gaussian $\chi^2$, and we must renormalize the result $\chi^2_{\rm fit}$ found for the fit by a factor ${\cal R}=1.110$ (1.027) for $\Delta_{\rm max}=6$ (9) to get the expected Gaussian result $\chi^2={\cal R}\times \chi^2_{\rm fit}$. This renormalized $\chi^2$ has the usual statistical interpretation.

Our original data set contained 167 datum points. In the analyses discussed in the next sections, we found the same 8 outlying points in fits performed with and without the high-energy constraints in \eq{HEconstraints}. Only 2.3 points with $\delchisqW>6$ would be expected for a Gaussian distribution of the data. The contribution of the outlying points to the total $\chi^2$ was essentially the same in the two cases. These outliers, if included, would increase the final $\chi^2$ of the fits by about 57\% relative to that of the points retained. For example, for the final 12 parameter fit using the high-energy constraints,  $\chi^2_{\rm fit}=161.2$ with an average $\chi^2$ per point of 1.01. The extra contribution of the outlying points in the original Lorentzian fit was 91.5, an average $\chi^2$ per point of 11.4 with actual values ranging from 6.6, slightly above the cutoff, to 28. We note finally that the outlying points are not concentrated in a way likely to affect our conclusions about high-energy scattering, with one point each in $\rho$ for $pp$ and $\bar{p}p$ scattering and three points in $B_{pp}$ distributed over the range $6.9\leq W\leq 62.5$ GeV, one in $\sigma_{\rm tot}^{\bar{p}p}$ at 8.76 GeV, one in  $\sigma_{\rm elas}^{\bar{p}p}$ at 900 GeV, and one in $\sigma_{\rm inel}^{pp}$ at 1800 GeV.

%%%%%%%%%%%%%%%%%%%%%%%

\subsection{Fit without high-energy constraints \label{subset:fit_LEconstraints}}

\begin{figure}[htbp]
\includegraphics{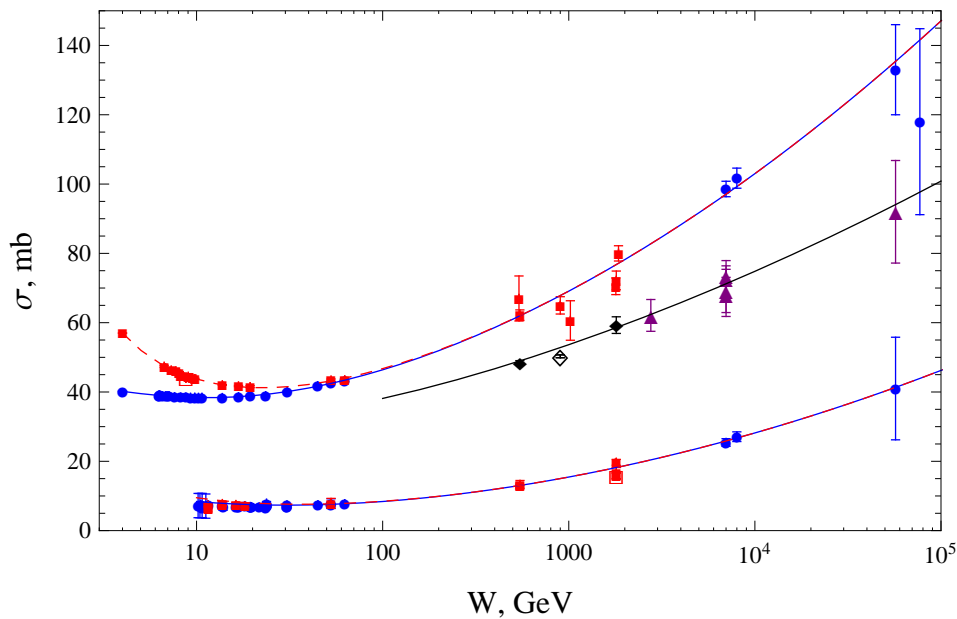}
\caption{Fits, top to bottom, to the total, inelastic, and elastic scattering cross sections  using the low-energy analyticity constraints in \eq{analyticity_const} and \eq{analyticity_const2} and the ratio constraints on the Regge-like contributions to the low-energy cross sections in \eq{ratio_const}: $\sigma_{\rm tot}^{\bar{p}p}$ and  $\sigma_{\rm elas}^{\bar{p}p}$ (red) squares and dashed (red) line; $\sigma_{\rm tot}^{pp}$  and $\sigma_{\rm elas}^{pp}$ (blue) dots and solid (blue) line; $\sigma_{\rm inel}^{\bar{p}p}$ (black) diamonds and line; $\sigma_{\rm inel}^{pp} $ (purple) triangles.  The fit used only data  on $\sigma_{\rm tot}$ for $W\geq 6$ GeV, $\sigma_{\rm elas}$ for $W\geq 30$ GeV, and $\sigma_{\rm inel}$ for $W\geq 540$ GeV.  The curve for $\sigma_{\rm elas}$ includes data down to 10 GeV to show how the cross section is tied down at lower energies.  Outlying points not used in the fit are shown with large open symbols surrounding the central points; the size of those symbols does not reflect the quoted errors of the measurement. }
\label{fig:xsectionsnoBD}
\end{figure}

We first consider the results of a global fit to the data on $\sigma_{\rm tot}$, $\sigma_{\rm elas}$, $\sigma_{\rm inel}$, $\rho$, and $B$ which is not constrained by the black-disk conditions in \eq{HEconstraints} at very high energies. We {\em did} use the low-energy constraints on the cross sections in   Eqs.\ (\ref{analyticity_const}) and (\ref{analyticity_const2}),  and the new ratio constraints on the coefficients of the Regge-like terms in \eq{ratio_const}; these constraints are essential in tying down the cross sections at low energies. The sieve algorithm was used to filter the data resulting in the elimination of  8 outliers among 167 datum points as noted above.  Combined plots of the cross sections  from the fit are shown in Fig. \ref{fig:xsectionsnoBD}. We do not show the fits to $\rho$ and $B$; the curves are nearly indistinguishable from those in \fig{fig:Brho} shown later.

%%%%%%%%%%%%%%%%%%%%%
%%%%%% FIG 1 cross sections no HE constraints  %%%%%

%%%%%%%%%%%%%%%%%%%%%
%

Table \ref{table1:fit14p} shows the results of this 14-parameter $\chi^2$ fit. As seen from the table, the raw $\chi^2$ per degree of freedom is 1.11, while the renormalized $\chi^2$ per degree of freedom is 1.23. This is a very good fit, especially considering the amount of data used.

%%%%%%%  Table 1 FINAL %%%%%%%%%%%%%%
%%%%%%%%%%%%%%%%%%%%%%%%%%%%%%

\begin{table}[h,t]                   % Use "table" environment, but also
				 % use  "tabular" environment below.
%
\def\arraystretch{1.15}            % Make the space between rows in the Table,
\begin{center}				  % 1.5 x bigger than the default spacing.
\begin{tabular}[b]{|l||c||}

\hline
{\rm Parameters} & $\delchimax=6$ \\
\hline
      $c_0$\ \ \   (mb) & $23.54 \pm 4.94$ \\
      $c_1$\ \ \   (mb) & $0.2043 \pm 1.023$ \\
      $c_2$\ \ \ \   (mb) & $0.2328 \pm 0.0381 $ \\
      $b_0$\ \ \   (mb) & $7.436 \pm 2.330 $ \\
      $b_1$\ \ \   (mb) & $-1.036 \pm 0.354$ \\
      $b_2$\ \ \   (mb) & $0.1230 \pm 0.015$ \\
      $a_0$\ \ \   (GeV$^{-2}$) & $10.38\pm 1.27$ \\
      $a_1$\ \ \   (GeV$^{-2}$) & $0.1304\pm 0.2190$ \\
      $a_2$\ \ \   (GeV$^{-2}$) & $0.02356\pm 0.0091$ \\
      $\beta$\ \   (mb) & $45.05\pm 6.42 $ \\
       $\beta_e$\ \   (mb) & $14.51\pm 2.07 $ \\
      $\beta_B$\ \   (GeV$^{-2}$) & $0.4634\pm 2.110 $ \\
      $f(0)$ (mb GeV) & $2.095\pm 0.569$ \\
      $\delta$\ \ \   (GeV$^{-2}$) & $-29.05 \pm 0.90 $ \\
      $\delta_e$\ \ \   (GeV$^{-2}$) & $-5.897\pm 0.182$ \\
      $\delta_B$\ \ \   (GeV$^{-2}$) & $-8.115\pm 0.513$ \\
      $\alpha$ & $0.4069\pm 0.006$ \\
      $\mu$ & $0.6593\pm 0.0449$ \\
\hline
\hline
	$\chi^2_{\rm min}$ & 160.864\\
	${\cal R}\times\chi^2_{\rm min}$ & 178.483\\
	Degrees of freedom (d.o.f). &145\\
\hline
	${\cal R}\times\chi^2_{\rm min}$/d.o.f.&1.231\\
\hline
\end{tabular}
     %\vspace{1in} \\
     \caption{\protect\small The results for our 14-parameter $\chi^2$ fit to the $\bar{p}p$ and $pp$ total, elastic, and inelastic cross sections, $\rho$ values and slope parameters $B$ using expressions in Eqs.\  (\ref{sigma0})--(\ref{B}), the low-energy constraints in Eqs.\  (\ref{analyticity_const}),  (\ref{analyticity_const2}), and (\ref{ratio_const}), and the cut $\delchimax=6$ in the sieve analysis of the data. The renormalized $\chi^2_{\rm min}$/d.o.f.,  taking into account the effects of the $\delchimax$ cut, is given in the row  labeled ${\cal R}\times\chi^2_{\rm min}$/d.o.f., with ${\cal R}(6)=1.110$.  \label{table1:fit14p}}
\end{center}
\end{table}
\def\arraystretch{1}  %Restore the default row spacing in the Table.
%%%
%%%%%%%%%%%%%%%%%%%%%%%%%%%%%%%%%%%%%%%

It is very interesting to use the results from this fit, constrained only at low energies, to examine the very-high-energy behavior projected for the cross sections and $B$.  We find from Table  \ref{table1:fit14p} that
\ba
\frac{\sigma_{\rm elas}}{ \sigma_{\rm tot}}\rightarrow \frac{b_2}{ c_2}=\frac{0.1230}{ 0.2328}=0.528 \pm 0.108,\quad {\rm as\ } s\rightarrow \infty.
\ea
The  deviation of this value of the ratio from  the expected value 1/2 for for black-disk scattering at infinity energy is well within the uncertainty of the fit.

We find that the ratio of the fitted value of the ratio of $B$ to its black-disk value $\sigma_{\rm tot}/8\pi$ also agrees very well with its expected value of 1 at high energies,
\be
(0.3894) {8 \pi} \frac{a_2}{c_2} =0.990 \pm 0.415,\quad  {\rm as\ } s\rightarrow \infty.
\label{bd2}
\ee

We conclude that these results, obtained from a fit  which used only the low-energy constraints in Eqs.\ (\ref{analyticity_const}), (\ref{analyticity_const2}), and (\ref{ratio_const}), give {\em strong} evidence both that  $pp$ and $\bar{p}p$ scattering can be described asymptotically as black-disk scattering, and that the limiting $\ln^2s$ behavior is already evident at present energies. The use of the constraints ties down the low-energy part of the fit, fixing the values of the total cross sections at 4 GeV and the ratios of the coefficients of the Regge-like terms in the cross sections. The low energy fit is excellent, and gives  slopes of the total cross sections with respect to $\nu/m$ at 4 GeV which agree reasonably well with those estimated from lower energy data \cite{blockrev} even though the data used in the fit was confined to energies above 6 GeV.

%%%%%%%%%%%%%%%%%%%%%%%

\subsection{Fit using the black disk constraints \label{subsec:black-disk}}

We have used the general parametrizations in Eqs.\ (\ref{sigma_tot})--(\ref{B}), with the low-energy constraints in Eqs.\ (\ref{analyticity_const}), (\ref{analyticity_const2}) and (\ref{ratio_const}), and the high-energy black-disk constraints \eq{HEconstraints} all imposed, to fit the combined $pp$ and $\bar{p}p$ data over the same energy ranges as above. The sieve algorithm was again used to eliminate the same 8 outliers among 167 datum points. There are now only 12 parameters.

The result of the fit is excellent as seen in the last lines in Table \ref{table2:fit12p}, with a $\chi^2$ of 161 for 147 degrees of freedom for a raw $\chi^2$ per d.o.f. of 1.10, and a renormalized $\chi^2/{\rm d.o.f.}$ of 1.22. As would be expected, the parameters of the fit  have smaller uncertainties than in the previous fit using only the low-energy constraints, and with the exception of $a_1$, change only within the previous uncertainties.

\begin{figure}[htbp]
\includegraphics{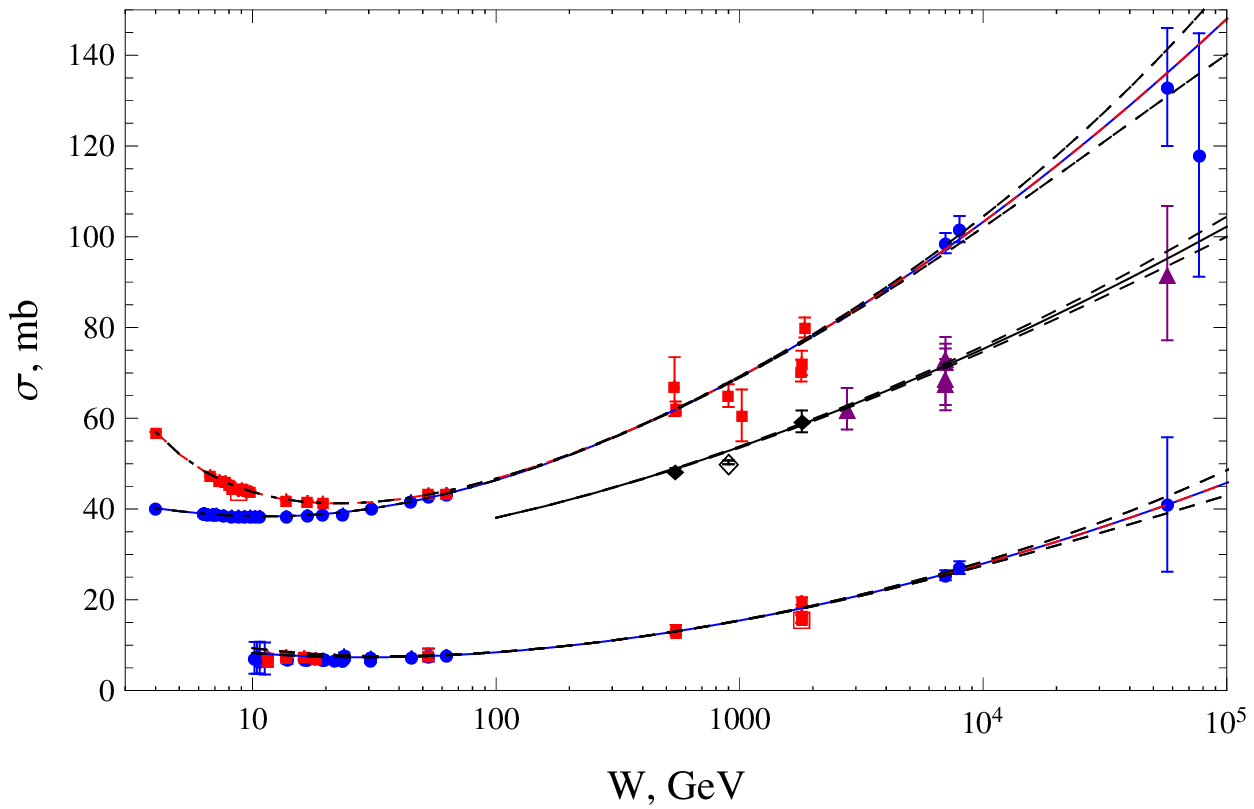}
\caption{Fits, top to bottom, to the total, inelastic, and elastic scattering cross sections  using high-energy black-disk constraints in \eq{HEconstraints} as well as the the low-energy analyticity constraints in \eq{analyticity_const} and \eq{analyticity_const2} and the ratio constraints on the Regge-like contributions to the low-energy cross sections in \eq{ratio_const}: $\sigma_{\rm tot}^{\bar{p}p}$ and  $\sigma_{\rm elas}^{\bar{p}p}$ (red) squares and dashed (red) line; $\sigma_{\rm tot}^{pp}$  and $\sigma_{\rm elas}^{pp}$ (blue) dots and solid (blue) line; $\sigma_{\rm inel}^{\bar{p}p}$ (black) diamonds and line; $\sigma_{\rm inel}^{pp} $  (purple) triangles.   The fit used only data  on $\sigma_{\rm tot}$ for $W\geq 6$ GeV, $\sigma_{\rm elas}$ for $W\geq 30$ GeV, and $\sigma_{\rm inel}$ for $W\geq 540$ GeV.  The curve for $\sigma_{\rm elas}$ includes data down to 10 GeV to show how the cross section is tied down at lower energies. Outlying points identified in the sieve analysis and not used in the fit are shown with large open symbols surrounding the central points; the size of those symbols is not connected to the quoted errors. The statistical error bands determined by the error analysis are shown.  }
\label{fig:xsectionsBD}
\end{figure}

\begin{figure}[htbp]
\includegraphics{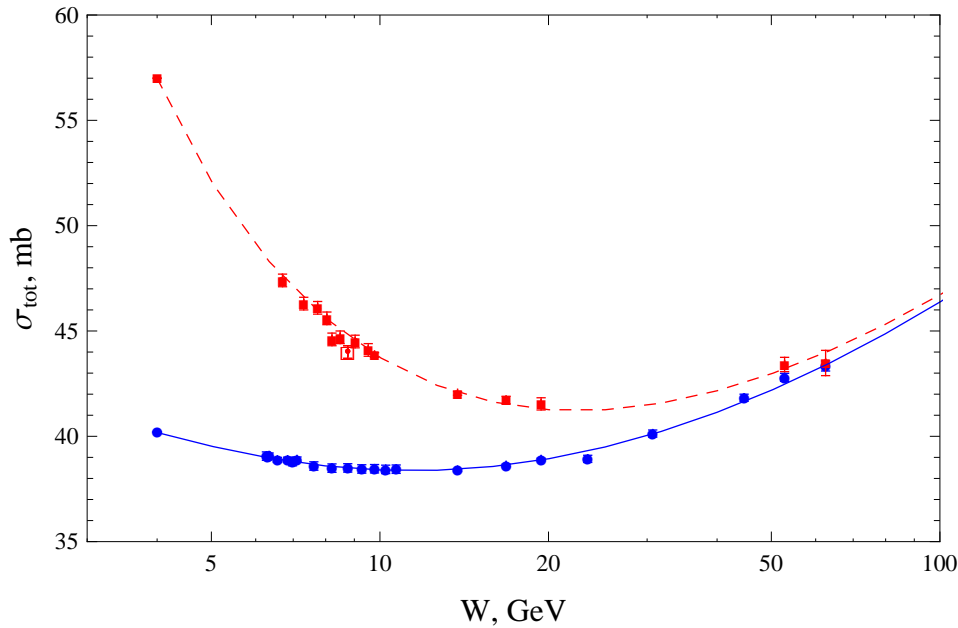}
\caption{Curves showing the fits to $\sigma_{\rm tot}^{pp}$, (blue) dots and solid (blue) line, and $\sigma_{\rm tot}^{\bar{p}p}$, (red) squares and dashed (red) line, at low energies, extending the curves for the total cross sections in \fig{fig:xsectionsBD}. The fits  used  the low-energy analyticity constraints in Eqs.\ (\ref{analyticity_const}) and (\ref{analyticity_const2}), the ratio constraints on the Regge-like contributions to the low-energy cross sections in \eq{ratio_const}, and the black-disk high-energy constraints in \eq{HEconstraints}. The $\bar{p}p$ outlier eliminated in the sieve analysis is shown with a large open symbol surrounding the central point; the size of the symbol does not reflect the quoted accuracy of the measured value. The fixed values of the cross sections at 4 GeV from the low-energy data are also shown.  }
\label{fig:xsectionsLE}
\end{figure}

\begin{figure}[htbp]
\includegraphics{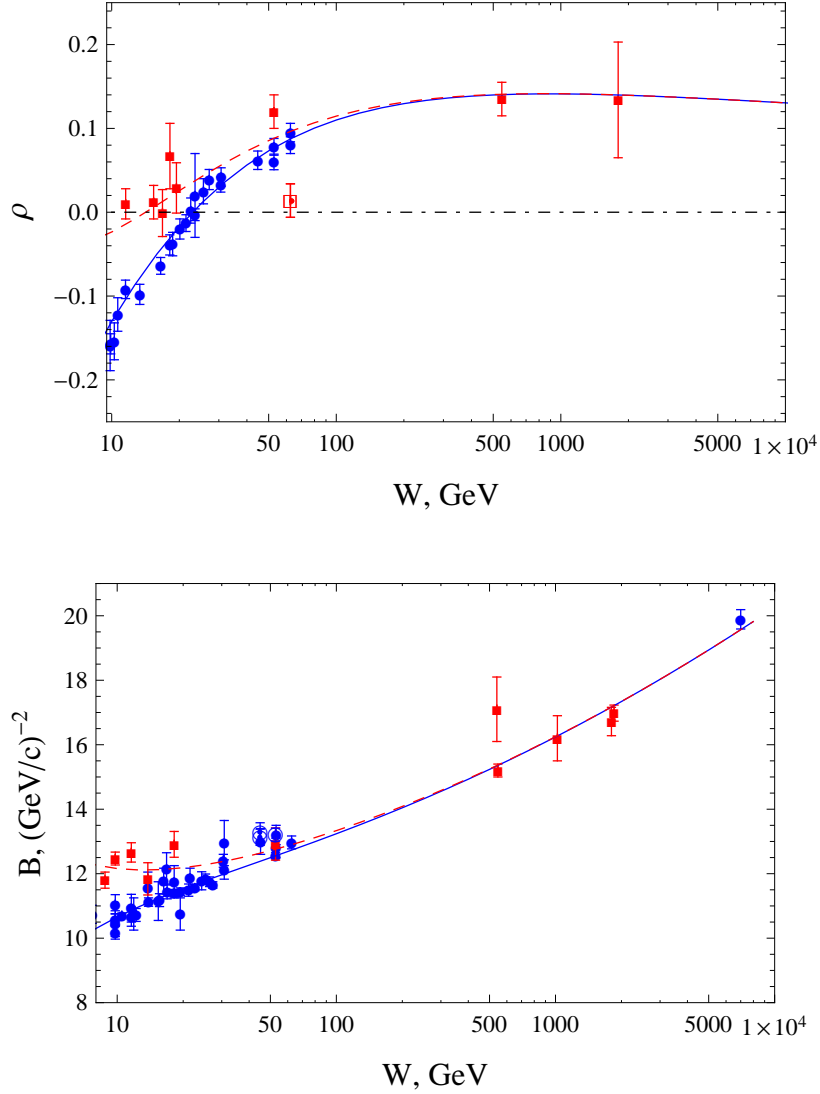}
 \caption{Top panel: fits to the ratios $\rho$ of the real to the imaginary parts of the forward scattering amplitudes for $pp$ (blue dots and solid blue line), and $\bar{p}p$ scattering (red squares and dashed red line). Lower panel: fits to the  logarithmic slope parameters for the elastic differential scattering cross sections $d\sigma/dt$ for $pp$ (blue dots and solid line) and $\bar{p}p$ (red squares and dashed line) scattering. The fits to $\rho$ and $B$ used only data above 6 GeV, and imposed  the low-energy constraints on the parameters in Eqs.\ (\ref{analyticity_const}), (\ref{analyticity_const2}), and (\ref{ratio_const}), and the high-energy asymptotic black-disk constraints in \eq{HEconstraints}. In both cases, the datum points eliminated in the sieve analysis are shown with large open symbols surrounding the central point; the size of the open symbols does not reflect the quoted accuracy of the measurement. The error bands estimated from the uncertainties in the parameters are too narrow to show in the figure. }
\label{fig:Brho}
\end{figure}

 We give  combined plots of the total, inelastic, and elastic cross sections at high energies in  Fig. \ref{fig:xsectionsBD} and show the lower-energy behavior of $\sigma_{\rm tot}$ in \fig{fig:xsectionsLE}. The fitted curves for $\rho$ and $B$ are compared with those data in Fig.\ \ref{fig:Brho}. All the data are shown, including the two cross section points, the two values of $\rho$, and the three values of $B$  dropped in the sieve analysis. We also show the statistical error bands for the fit; these show that the fit  is very tightly constrained over the region of the data. The consistency with the fit without the high-energy constraints and the rather small 11\% uncertainty in $c_2= 0.2425 \pm 0.0268$ mb indicate that the asymptotic cross sections are also well-determined.

 As shown in \fig{fig:xsectionsLE}, we fit the total cross sections  very well  at energies down the 6 GeV,  the lower limit used in our analysis. The curves match the data  and extend smoothly to the fixed  values at 4 GeV used in the low-energy constrains in \eq{analyticity_const} and \eq{analyticity_const2}. Even though the slopes $d\sigma_{\rm tot}/d(\nu/m)=(m^2/W)d\sigma_{\rm tot}/dW$ at $\nu_0 =7.59$ GeV or $W_0=4$ GeV were not used in the fitting by imposing the second set of analyticity constraints in \cite{block_analytic,blockhalzenfit}, the calculated slopes, respectively -1.38\,(-0.169) mb for $\bar{p}p$\,($pp$),  match well with the slopes -1.45\,(-0.231) determined  from the dense data around 4 GeV \cite{blockrev}.

 The present fits agree well with those of earlier work based on more limited data. The results of Block and Halzen \cite{blockhalzenfit,blockhalzen2} used only the total cross sections and $\rho$ values up to 1.8 TeV, without including the elastic or inelastic cross sections or measured values of $B$. Their results gave $c_2=0.2817\pm 0.0064$ mb and predicted total cross sections of $95.4\pm 1.1$ mb, $97.6\pm 1.1$ mb, and $134.8\pm 4.5$ mb at $W=7$, 8, and 57 TeV, in substantial agreement with the values $98.6\pm 2.2$, $101\pm 2.1$ mb, and $133\pm 13)\, ({\rm stat})+17 (-20) \,({\rm sys})\pm 16\, ({\rm Glauber})$) mb found by TOTEM \cite{totem2013,totem2015} and AUGER \cite{POAp-air}.

 Our results for the completely constrained fit using the total, elastic, and inelastic cross sections, $\rho$, and $B$ give $c_2=0.2425\pm0.0268$ mb, $\sigma_{\rm tot}= 97.27\pm 0.86$ mb and $99.49\pm 0.97$ mb at 7 and 8 TeV, and $136.1\pm 5.2$ mb at 57 TeV. We conclude that the fits are consistent and stable. An important reason for this stability is our imposition of the low-energy constraints: the non-leading terms in the parametrizations in Eqs.\ (\ref{sigma0})-(\ref{B}) are less well determined if the constraints are ignored, indirectly affecting the high-energy terms and the asymptotic behavior for $W$ large.

%%%%%%%%%%%%%%%%%%%%%
%%%%%% FIG 2 cross sections with black disk constraints  %%%%%
%%%%%%%%%%%%%%%%%%%%%
%

%%%%%%%%%%%%%%%%%%%%%
%%%%%% FIG 3 LE cross sections with black disk constraints  %%%%%
%%%%%%%%%%%%%%%%%%%%%
%

%%%%%%%%%%%%%%%%%%%%%
%%%%%% FIG 4 rho, B BDconstraints  %%%%%
%%%%%%%%%%%%%%%%%%%%%
%

%%%%%%%%%  Table 2 FINAL %%%%%%%%%%%%%%%%%
%%%%%%%%%%%%%%%%%%%%%%%%%%%%%%%%%%

\begin{table}[h,t]                   % Use "table" environment, but also
				 % use  "tabular" environment below.
%
\def\arraystretch{1.15}            % Make the space between rows in the Table,
\begin{center}				  % 1.5 x bigger than the default spacing.
\begin{tabular}[b]{|l||c||}

\hline
{\rm Parameters} & $\delchimax=6$ \\
\hline
       $c_0$\ \ \   (mb) & $26.76 \pm 3.49$ \\
      $c_1$\ \ \   (mb) & $-0.049 \pm 0.715$ \\
      $c_2$\ \ \ \   (mb) & $0.2425 \pm 0.0268 $ \\
      $b_0$\ \ \   (mb) & $7.565 \pm 2.011 $ \\
      $b_1$\ \ \   (mb) & $-1.022 \pm 0.322$ \\
      $b_2$\ \ \    (mb) & $0.1213 \pm  0.0134$ \\
      $a_0$\ \ \   (GeV$^{-2}$) & $10.55 \pm 0.44$ \\
      $a_1$\ \ \   (GeV$^{-2}$) & $1.013 \pm 0.069$ \\
      $a_2$\ \ \   GeV$^{-2}$) & $0.02478 \pm 0.0027$\\
      $\beta$\ \   (mb) & $43.49 \pm 3.49 $ \\
      $\beta_e$\ \  (mb) & $14.00 \pm 1.12 $ \\
      $\beta_B$\ \   (GeV$^{-2}$) & $-0.1632 \pm 0.8759 $ \\
      $f(0)$ (mb GeV) & $2.137 \pm 0.561$\\
      $\delta$\ \ (mb) & $-29.05 \pm 0.90 $\\
      $\delta_e$\ \  (mb) & $-5.897 \pm 0.182 $ \\
      $\delta_B$\ \ \   (GeV$^{-2}$) & $-8.157 \pm 0.511$ \\
      $\alpha$ & $0.4068 \pm 0.0060$ \\
      $\mu $ & $0.6486 \pm 0.0353$ \\
\hline
\hline
	$\chi^2_{\rm min}$&161.15 \\
	${\cal R}\times\chi^2_{\rm min}$ & 178.80\\
	Degrees of freedom (d.o.f).&147\\
\hline
	${\cal R}\times\chi^2_{\rm min}$/d.o.f.&1.216\\
\hline
\end{tabular}
     %\vspace{1in} \\
     \caption{\protect\small The results for our 12-parameter $\chi^2$ fit to the $\bar{p}p$ and $pp$ total, elastic, and inelastic cross sections, $\rho$ values and slope parameters $B$ using expressions in Eqs.\  (\ref{sigma0})--(\ref{B}), the low-energy constraints in Eqs.\  (\ref{analyticity_const}),  (\ref{analyticity_const2}), and (\ref{ratio_const}), the black-disk constraints in \eq{HEconstraints}, and  the cut $\delchimax=6$ in the sieve filtering of the data which eliminated 8 outlying points. The renormalized $\chi^2_{\rm min}$/d.o.f.,  taking into account the effects of the $\delchimax$ cut, is given in the row  labeled ${\cal R}\times\chi^2_{\rm min}$/d.o.f., with ${\cal R}(6)=1.110$.   \label{table2:fit12p}}
\end{center}
\end{table}
\def\arraystretch{1}  %Restore the default row spacing in the Table.
%%%%%%%%%%%%%%%%%%%%%%%%%%%%%%%%%%%%%
%%%%%%%%%%%%%%%%%%%%%%%%%%%%%%%%%%%%%%%

The crossing-even high energy inelastic cross section $\sigma^0_{\rm inel}(\nu)$, valid in the energy domain $\sqrt s \ge 100$ GeV where the odd Regge-like terms are very small and $\sigma_{\rm tot}^{pp}$ and $\sigma_{\rm tot }^{\bar{p}p}$ are essentially equal, is given by
\ba
\sigma_{\rm inel}^0(\nu)&=& (19.20\pm 4.03)+(0.9729\pm 0.784) \ln\left(\frac{\nu}{m}\right)\nonumber\\
&&+ (0.1212\pm 0.0300) \ln^2\left(\frac{\nu}{m}\right)+(29.49\pm 3.66)\left(\frac{\nu}{m}\right)^{-0.3514} \ {\rm mb},
\label{finalinelastic}
\ea
%.
the difference of the expressions for $\sigma_{\rm tot}$ and $\sigma_{\rm elas}$ with the coefficients in Table\ \ref{table2:fit12p}.

For the convenience of the reader, we give the numerical predictions from the fit for the high energy $pp$ (or $\bar{p}p$) total, inelastic, and elastic cross sections,  $\rho$, and $B$   in Table \ref{table:predictions}.

%%%%%% Table 3 PREDICTIONS %%%%%%%%%%%
%%%%%%%%%%%%%%%%%%%%%%%%%%%%%

\begin{table}[h,t]                   % Use "table" environment, but also
				 % use  "tabular" environment below.
\def\arraystretch{1.5}            % Make the space between rows in the Table,
				  % 1.5 x bigger than the default spacing.
%
\begin{tabular}[b]{|l||c|c|c|c|c||}
    \cline{1-6}
      \multicolumn{1}{|l||}{ $\sqrt s$ (GeV)}
      &\multicolumn{1}{c|}{$\sigma_{\rm tot,pp}$ (mb)}
      &\multicolumn{1}{c|}{$\sigma_{\rm inel,pp}$ (mb)}&\multicolumn{1}{c|}{$\sigma_{\rm elas,pp}$ (mb)}&\multicolumn{1}{c|}{$\rho_{\rm pp}$}&\multicolumn{1}{c||}{$B_{\rm pp}$ (GeV/c)$^{-2}$}\\

      \hline\hline
	 540&$61.81\pm0.10$&$48.83\pm0.10$&$12.99\pm0.03$&$0.140\pm0.000$&$15.34\pm0.01$\\\hline
    900&$67.78\pm0.15$&$52.80\pm0.15$&$14.99\pm0.05$&$0.141\pm0.000$&$16.08\pm0.01$\\\hline
  1,800&$76.78\pm0.26$&$58.65\pm0.24$&$18.13\pm0.09$&$0.140\pm0.000$&$17.17\pm0.02$\\\hline
  7,000&$97.27\pm0.86$&$71.57\pm0.52$&$25.70\pm0.32$&$0.133\pm0.000$&$19.57\pm0.04$\\\hline
  8,000&$99.49\pm0.97$&$72.94\pm0.56$&$26.54\pm0.36$&$0.132\pm0.000$&$19.82\pm0.04$\\\hline
 13,000&$107.8\pm1.5$&$78.08\pm0.72$&$29.75\pm0.56$&$0.129\pm0.000$&$20.78\pm0.05$\\\hline
 14,000&$109.2\pm1.6$&$78.89\pm0.75$&$30.26\pm0.60$&$0.128\pm0.000$&$20.93\pm0.06$\\\hline
 57,000&$136.1\pm5.2$&$95.16\pm1.63$&$40.95\pm1.87$&$0.119\pm0.000$&$23.99\pm0.127$\\\hline
100,000&$148.0\pm7.8$&$102.2\pm2.21$&$45.77\pm2.77$&$0.115\pm0.000$&$25.32\pm0.173$\\\hline

\end{tabular}
     %\vspace{1in} \\
     \caption{\protect\small Predictions of high energy $pp$  total, inelastic, and elastic cross sections, $\rho$-values and $B$, using the parameters of Table \ref{table2:fit12p} in the expressions in Eqs.\ (\ref{sigma0})--(\ref{B}).\label{table:predictions}
}
%     \\
\end{table}
%%%%%%%%%%%%%%%%%%%%%%%%%%%%%%
%%%%%%%%%%%%%%%%%%%%%%%%%%%%%%

\begin{figure}[htbp]
\includegraphics{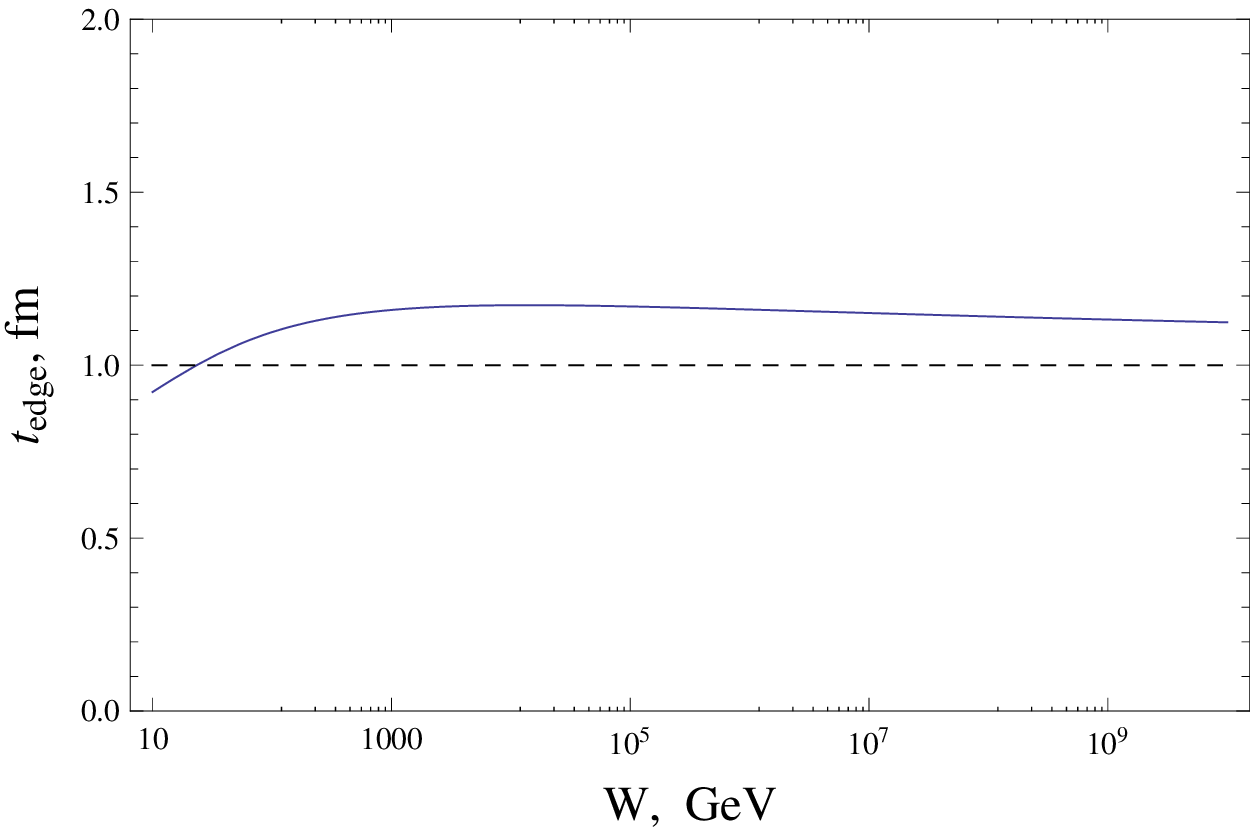}
 \caption{Solid curve: plot of the width $t_{\rm edge}$ of the soft edge in the crossing-even part of the $pp$ and $\bar{p}p$ scattering amplitudes as a function of energy from 10 to $10^{10}$ GeV. The horizontal dashed line is a $t_{\rm edge}=1$ fm.  }
 \label{fig:edge}
\end{figure}

We remark finally that, although the $pp$ and $\bar{p}p$ scattering amplitudes approach the black-disk limit at very high energies in the sense that $\sigma_{\rm elas}/\sigma_{\rm tot}\rightarrow 1/2$  and $B\rightarrow \sigma_{\rm tot}/8\pi$, there is not a sharp cutoff in those distributions in impact parameter space as in the classic black-disk model with unit amplitudes for $b<R$ and zero amplitudes for $b>R$, $R=\sqrt{\sigma_{\rm tot}/2\pi}$. Rather, as observed in \cite{edge} and studied in detail in \cite{bdhh_eikonal}, the scattering amplitudes have a smooth edge region of  approximately constant width $t_{\rm edge}\approx 1$ fm in impact parameter space, with
\be
\label{edge}
t_{\rm edge} \approx (2\sigma_{\rm inel}-\sigma_{\rm tot})/\sqrt{\pi\sigma_{\rm tot}/2}.
\ee
We show this in \fig{fig:edge} using the parameters in Table \ref{table2:fit12p} for the fit with the black-disk constraints imposed. Given the accuracy of the fit, we conclude that there is no evidence in the present data that the edge width shrinks significantly at very high energies, with $t_{\rm edge}\rightarrow 1.018$ fm for $s\rightarrow\infty$.

%%%%%%%%%% FIG. 5 Edge  %%%%%%%%%%%%%
%%%%%%%%%%%%%%%%%%%%%%%%%%%%%%
%%%%%%%%%%%%%%%%%%%%%%
%%%%%%%%%%%%%%%%%%%%%%

%%%%%%%%%%%%%%%%%%%%%%%
%%%%%%%%%%%%%%%%%%%%%%%

\section{Conclusions \label{sec:conclusions}}

 We have shown that we can obtain a very good fit to all the high-energy data on the  total, elastic, and inelastic $pp$ and $\bar{p}p$ scattering  cross sections, the ratios $\rho$ of the real to the imaginary parts of the forward scattering amplitudes, and the logarithmic slopes $B$ of the elastic scattering cross sections,  using expressions quadratic in $\ln{s}$ with added falling Regge-like terms at low energies. The use of these expressions, introduced in \cite{blockcahn} on the basis of the Froissart bound, was justified  in \cite{bdhh_eikonal} for detailed eikonal descriptions of the scattering in which the eikonal function grows as a power of $s$. The Froissart bound is satisfied but is not an input in that analysis, nor is it directly a motivation for the forms chosen here for the cross sections, $\rho$, and $B$ in Eqs.\ (\ref{sigma0})--(\ref{B}).

 The initial fit we presented here used constraints on the values of the cross sections at $W=4$ GeV, and new relations for the ratios of coefficients of the the Regge-like terms in the cross sections, to fix the fit at low energies. The results show that the cross sections and values of $B$ obtained using the present data satisfy the conditions $\sigma_{\rm elas}/\sigma_{\rm tot}\rightarrow 1/2$ and $B\rightarrow \sigma_{\rm tot}/8\pi$ expected for black-disk scattering within the uncertainties in the fit. We regard these results as, first, a demonstration that  data at the energies currently accessible already  reflect the asymptotic $\ln^2s$ behavior of the cross sections, and second, as  convincing evidence  for black-disk behavior of the $pp$ and $\bar{p}p$ scattering amplitudes at very high energies.

We then presented a second fit in which we imposed the black-disk behavior as a constraint at high energies. This gives nearly identical results,  provides predictions for the cross sections at  energies higher than those accessible now, and sharpens the analysis of results on the soft edge region in the scattering amplitudes discussed earlier \cite{edge,bdhh_eikonal}.

It is known from the proton structure functions of deep inelastic scattering, and theoretically, that the proton interactions at high energies are determined mainly by the gluonic and associated flavor-independent sea quark structure of the proton.   We expect the same asymptotic structure for other hadrons, with a universal color confinement volume, implying that all hadronic cross sections, {\em e.g.}, the  $\pi^{\pm} p$ and $K^{\pm}p$ cross sections, should approach the same black-disk limit as found for the $pp$ and $\bar{p}p$ cross sections.  This picture is supported by the analysis of Ishida and Barger \cite{IshidaBarger} who fit the $\pi^{\pm}p$ and $K^{\pm}p$ cross sections and $\rho$ values using a parametrization equivalent to that used here, and with the fitted cross sections similarly constrained to agree with the low-energy data through continuous moment sum rules. Their results and those here are consistent with the existence of a universal black-disk limit. For extensive references on the possible theoretical origin of the universality, beginning with  L.L. Jenkovszky, B.V. Struminsky and A.N. Vall \cite{jenkovszky}, see \cite{IshidaBarger,giordano}.

These results could be modified with the advent of new physics at  higher energies which significantly changes the nature of the hadronic interactions. There is no evidence of such changes in the present scattering data.

%%%%%%%%%%%%%%%%%%%%%
%%%%%%%%%%%%%%%%%%%%%%

\begin{acknowledgments}

M.M.B., L.D., and F.H.\  would  like to thank the Aspen Center for Physics for its hospitality and for its partial support of this work under NSF Grant No. 1066293. F.H.'s research was supported in part by the U.S. National Science
Foundation under Grants No.~OPP-0236449 and PHY-0969061 and by the
University of Wisconsin Research Committee with funds granted by the Wisconsin Alumni Research
Foundation.   P.H.\ would like to thank Towson University Fisher College of Science and Mathematics for support.

\end{acknowledgments}

%%%%%%%%%%%%%%%%%%%%%%
%%%%%%%%%%%%%%%%%%%%%%

\bibliography{small_x_references}

\end{document}